\documentstyle[times,epsfig,prl,aps,amssymb]{revtex}
\begin{document}
\draft
\title{System Size Coherence Resonance}
\author{Ra\'ul Toral$^{1,2}$ \and Claudio R. Mirasso$^{1}$ \and
\and James D. Gunton$^{2,3}$}
\address{
(1)Departament de F{\'\i}sica, Universitat de les Illes Balears, E-07071 Palma de Mallorca, Spain.\\
(2)Instituto Mediterr\'aneo de Estudios Avanzados (IMEDEA),
CSIC-UIB, E-07071 Palma de Mallorca, Spain.\\ 
(3)Department of Physics, Lehigh University, Bethlehem, PA 18015, USA\footnote{Permanent Address}.\\
}
\maketitle

\begin{abstract} 

We show the existence of a system size coherence resonance effect for an
ensemble of globally coupled excitable systems. Namely, we demonstrate
numerically that the regularity in the signal emitted by an ensemble of
globally coupled Fitzhugh-Nagumo systems, under excitation by independent noise
sources, is optimal for a particular value of the number of coupled systems.
This resonance is shown through several different dynamical measures: the time
correlation function, correlation time and jitter. 

\end{abstract}

Noise induced resonance is a topic that has attracted a lot of attention in the
last years. In particular, it has been unambiguously shown that the response of
some systems to an external perturbation can be {\sl enhanced} by the presence
of noise ({\sl stochastic resonance}\cite{BSV81,NN81,JSP70,GHJM98}). A
different effect is that of {\sl coherence resonance}\cite{PK97} in which an
excitable system shows a maximum degree of regularity in the emitted signal in
the presence of the right amount of fluctuations (or the related one of
stochastic resonance without the need of an external forcing\cite{GDNH93,RS94})
Coherence resonance has also been studied in dynamical systems close to the
onset of a bifurcation\cite{NSS97}, as well as in other bistable and
oscillatory systems\cite{LS00,POC99}. It has also been analyzed in different
neuronal models such as the FitzHugh--Nagumo\cite{LS99,RP99} and
Hodgkin-Huxley\cite{LNK98} models. It has been observed experimentally in
electronic circuits, either excitable\cite{PHYS99,HYPS99} or
chaotic\cite{PTMCG01,CMT01}, and in lasers operating in an excitable
regime\cite{GGBT00}.

In an important recent paper \cite{PZC02}, Pikovsky {\sl et al.} have shown
that when one considers an ensemble of coupled bistable systems subjected to an
external periodic forcing (and in the presence of a constant amount of noise),
it turns out that an optimal response is obtained for an appropriate value of
the number $N$ of coupled systems. In other words, that there is a resonance
with respect to the \textbf{number} of coupled elements, rather to the usual one
that involves the noise level. The authors speculate that this system size
resonance might be relevant to neuronal dynamics, in which the neuronal
connections or the coupling strengths between neurons can be tuned in order to
achieve maximum sensitivity to external signals.

In this paper, we extend the previous result by considering an ensemble
of globally coupled excitable systems, each one under the influence of its
own noise with a fixed intensity, but without an external forcing. We show
that there is a coherence resonance effect as a function of the number $N$
of coupled systems. More specifically, we show that the excitable systems
pulse on average with a regularity which is optimal for a specific value of $N$.

Motivated by the biological applications suggested in \cite{PZC02}
we consider an ensemble of $i=1,\dots,N$ coupled FitzHugh--Nagumo
systems, each one described by the activator, $x_i$, and
inhibitor, $y_i$, variables. The FitzHugh--Nagumo model provides
the simplest representation of firing dynamics and has been widely
used as a prototypic model for spiking neurons as well as for
cardiac cells\cite{K99,GHM91}. The dynamical FitzHugh--Nagumo equations modified to account for the global coupling are as
follows:
\begin{eqnarray}
\label{eq:x}
\epsilon\dot{x_{i}}& = & x_{i}-\frac{1}{3}x_{i}^3-y_{i}+\frac{K}{N}\sum_{j=1}^N
(x_{j}-x_{i})\\
\label{eq:y}
\dot{y_{i}}& = & x_{i}+a+D\xi_{i}(t)
\end{eqnarray}
where independent noises of intensity $D$ have been added to the
slow variables $y_i$ as in ref. \cite{PK97}. The $\xi_i(t)$ are white noises with
Gaussian distribution of zero mean and correlations $\langle
\xi_i(t)\xi_j(t')\rangle = \delta_{ij}\delta(t-t')$. The
difference in the time scales of $x_i$ and $y_i$ is measured by
$\epsilon$, a small number. The systems are globally coupled, as
indicated by the last term of Eq.(\ref{eq:x}), where $K$ is the
coupling strength.

We consider the excitable regime, $a>1$. In the absence of coupling, $K=0$,
each FitzHugh--Nagumo system emits pulses. The pulses are trajectories that
exit the basin of attraction of the stable fixed point and are triggered by the
influence of the noise term. For zero coupling, coherence resonance exists when
the regularity of the time between pulses is optimal for a certain value of the
noise intensity $D$\cite{PK97}. This behavior is the same for all the systems
although the response of each one is uncorrelated with any other.

Let us now study the collective response of the coupled system and compare it with the individual responses. For the collective response, we
introduce the average values of the activator and inhibitor variables as
\begin{equation}
X(t)=\frac{1}{N}\sum_{i=1}^N x_i(t) \hspace{3.0cm} Y(t)=\frac{1}{N}\sum_{i=1}^N y_i(t)
\end{equation}
By following the approach by Desai and Zwanzig\cite{DZ78} (see also reference \cite{PZC02})
it is possible to reach an approximate effective equation for these average values of the form:
\begin{eqnarray}
\label{eq:bigx}
\epsilon \dot{X}=F(X)- Y
\\
\label{eq:bigy}
\dot{Y}=X+a+\frac{D}{\sqrt{N}}\xi(t)
\end{eqnarray}
where $\xi(t)$ is a white noise source. Although the exact form of the function
$F(X)$ and the analysis of the approximations assumed in the derivation will be presented elsewhere, we need only to remark here that in the (exact) equation for $Y(t)$ the noise intensity appears rescaled as $D/\sqrt{N}$. Therefore, this approximation suggests that the optimal effective noise intensity for the appearance of coherence resonance can be achieved by varying the number of coupled elements $N$, as in the case of stochastic resonance for the bistable system considered in \cite{PZC02}. To go beyond this approximation, we numerically integrate the equations of motion (1) and (2). 

\begin{figure}
\epsfxsize=0.75\textwidth\epsfysize=0.75\textwidth\centerline{\epsfbox{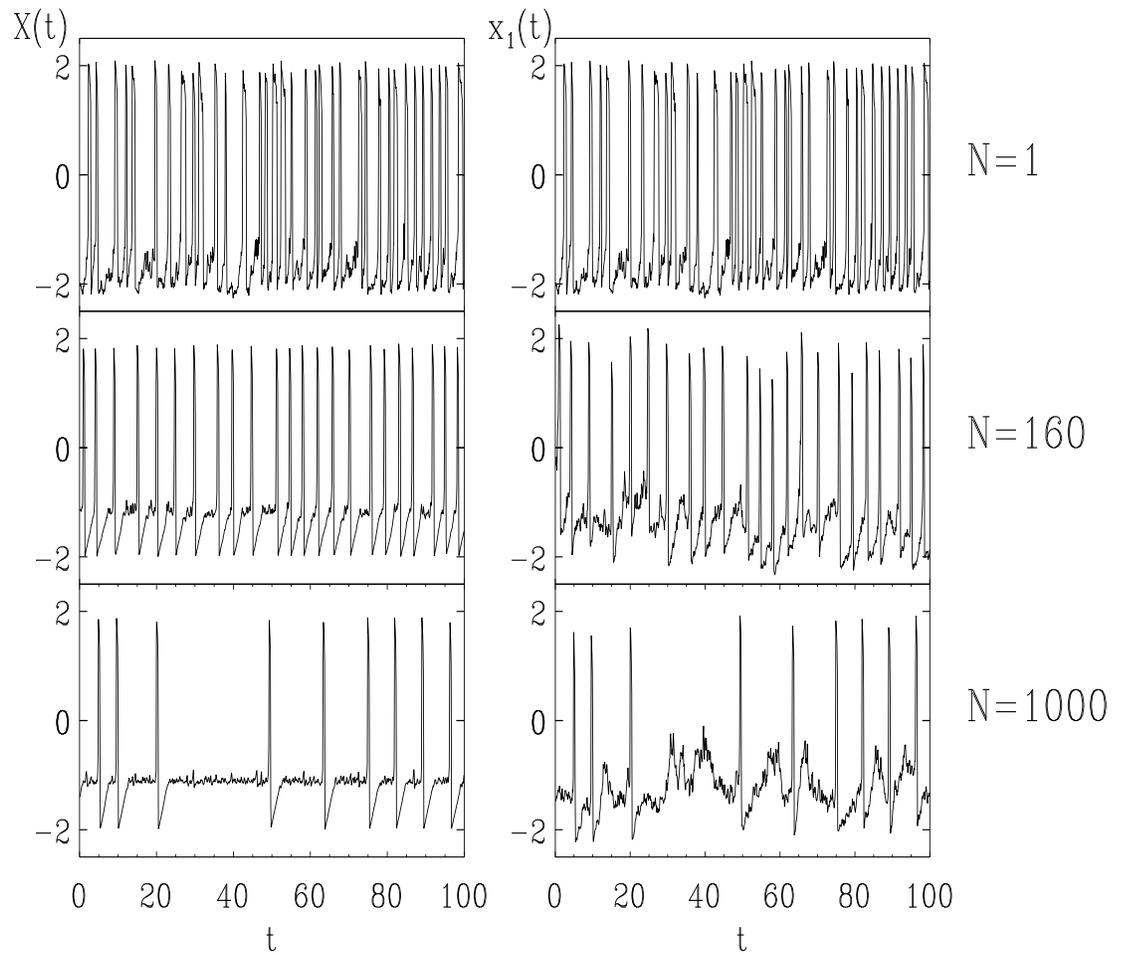}}
\caption{\label{fig1} Time series for the averaged variable $X(t)$ (left
panel), and for the individual variable $x_1(t)$ (right panel) of the set of coupled FitzHugh--Nagumo systems, as obtained from a numerical integration of
Eqs.(\ref{eq:x}-\ref{eq:y}), for different values of the number of coupled
elements: $N=1$ (top), $N=160$ (middle) and $N=1000$ (bottom). Observe that the
largest regularity is obtained for the intermediate value of $N$. The equations
have been integrated numerically using a stochastic Runge--Kutta method (known
as the Heun method [23]) with a time step $h=10^{-4}$ and setting the
following parameters: $a=1.1$, $\epsilon=0.01$, $K=2$, $D=0.7$.}  
\end{figure}

\begin{figure}
\epsfxsize=0.75\textwidth\epsfysize=0.75\textwidth\centerline{\epsfbox{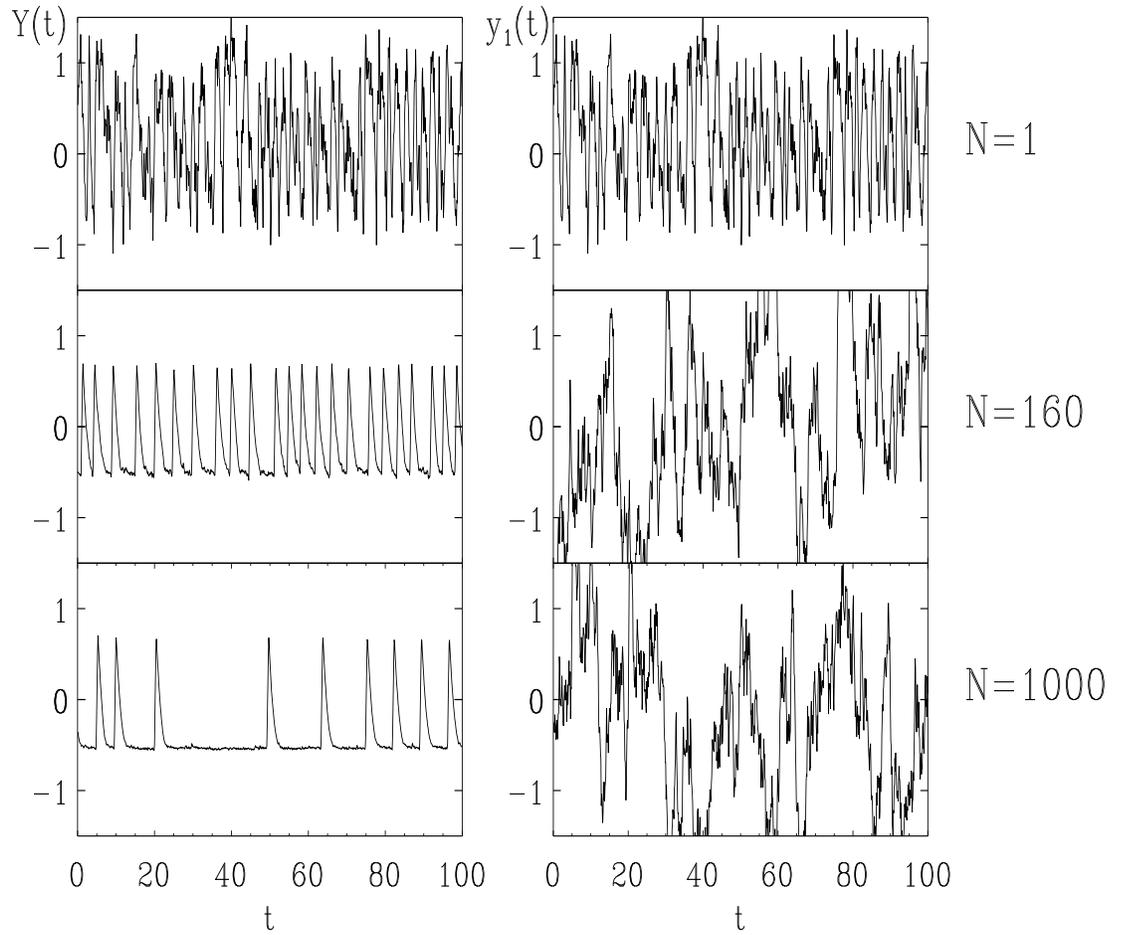}}
\caption{\label{fig2} Time series for the averaged variable $Y(t)$ (left panel), and the individual variable $y_1(t)$ (right panel) of the set of FitzHugh--Nagumo systems, Eqs.(\ref{eq:x}-\ref{eq:y}). Similarly as in Figure \ref{fig1}, observe
that again the largest regularity for the averaged $Y$ variable is obtained for the intermediate value of $N$. In this case, however, there is no obvious increase in the regularity of the $y_i$ individual variables.}
\end{figure}

Figure \ref{fig1} (left panel) shows the time trace for the variable $X(t)$
while \ref{fig1} (right panel) shows the time trace for the variable $x_i(t)$
of one of the elements chosen randomly, for three different values of the
number of coupled elements (see the caption of the figure for details of the
parameters). Notice that for $N=160$ the regularity of the emitted pulses is
better than that corresponding to larger or smaller values of $N$. This is a
clear signature of coherence resonance. Moreover, it can be seen that the
regularity in the averaged variable $X(t)$ is better than in one of the
individual elements, showing that the coupling allows for a smoothness of the
trace. It is worth noting that the peaks in the collective variable $X(t)$ and
$x_i(t)$ are very well synchronized in time indicating that the individual
systems are pulsing synchronously in time. In Figure \ref{fig2} (left panel) we
plot the time trace for the slow variable $Y(t)$, as well as a time trace for a
single $y_i(t)$ (right panel). At variance with the fast variable $X$, it turns
out that the averaged $Y(t)$ shows a very nice regular behavior for an
intermediate number of elements while the individual traces $y_i(t)$ do not. 

\begin{figure}
\epsfxsize=0.85\textwidth\epsfysize=0.5\textwidth\centerline{\epsfbox{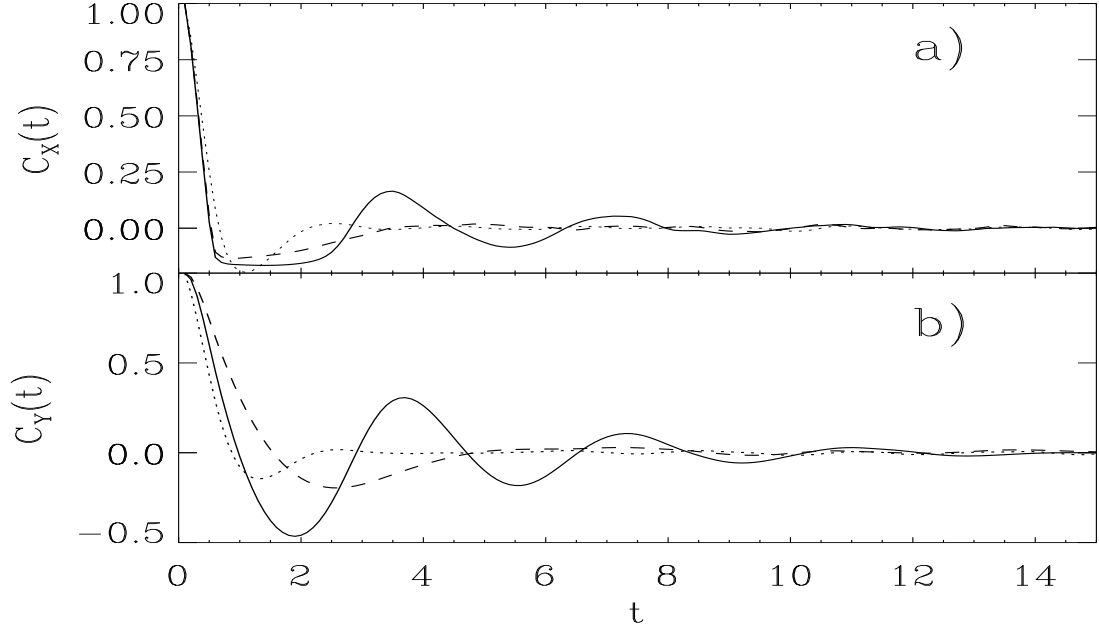}}
\caption{\label{fig3} Correlation functions $C_X(t)$ and $C_Y(t)$ of the
averaged variables $X(t)$ and $Y(t)$, respectively, for the cases of $N=1$
(dotted line), $N=160$ (solid line) and $N=1000$ (dashed line). Notice that, in
agreement with the qualitative results derived from figures \ref{fig1} and
\ref{fig2}, the slower decay of the correlations corresponds to the intermediate
values of the system size $N$. Same parameters as in figure \ref{fig1}.}
\end{figure}

\begin{figure}
\epsfxsize=0.85\textwidth\epsfysize=0.5\textwidth\centerline{\epsfbox{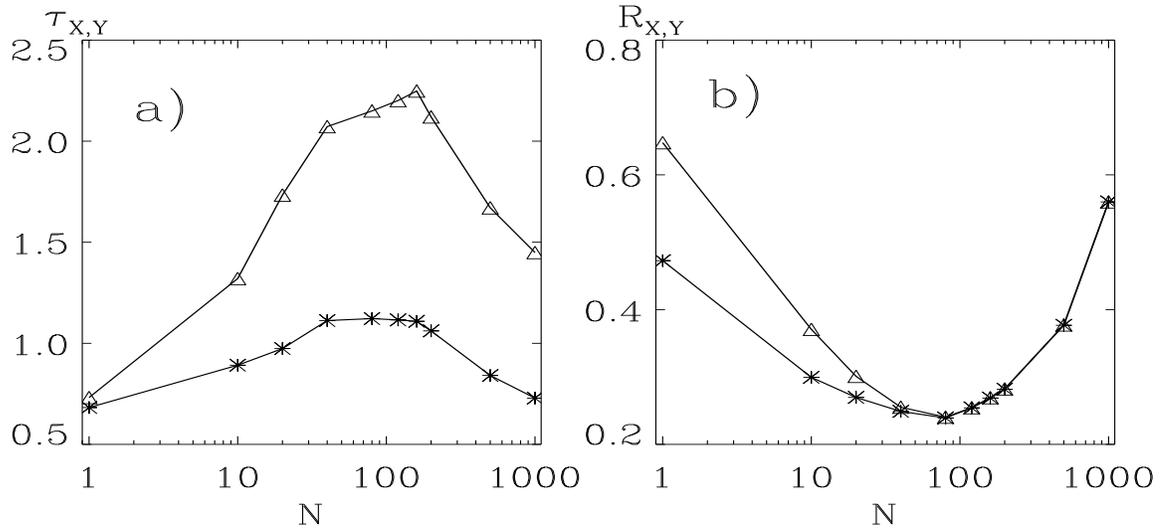}}
\caption{\label{fig4} Panel (a) plots the correlation times $\tau_X$ and $\tau_Y$ as obtained by integration of the absolute value of the respective correlation functions. Clear maxima (maximum extent of the correlations) can be observed around $N=160$. Panel (b) plots the jitter of the time between consecutive pulses of the collective variables $X(t)$ (stars) and $Y(t)$ (triangles). Clear minima (optimal regularity in the emitted pulses) can be observed around $N=80$ in both cases.}
\end{figure}

We have computed two indicators commonly used to quantify this
effect\cite{PK97}. First, we have computed the time correlation function $C_X(t)$ of the averaged $X$ variable, defined as
\begin{equation}
C_X(t)=\frac{\langle \delta X(t')\delta X(t+t')\rangle}{\langle \delta X(t')^2\rangle} \hspace{1.0cm} \delta X(t)=X(t)-\langle X(t')\rangle
\end{equation}
and similarly for the correlation function $C_Y(t)$ for the averaged $Y$
variable. Here the averages $\langle \,\rangle$ are with respect to the time
$t'$, after a small transient has been neglected. Figure \ref{fig3} shows this
correlation function for both the $X$ and $Y$ variables. It can be seen that
the correlations extend further in time for an intermediate value, neither very
large nor very small, of the number of coupled systems $N$. To obtain a quantitative indicator of this effect, we define the
characteristic correlation times $\tau_X$ and $\tau_Y$ for each variable as
\begin{equation}
\tau_{X,Y}=\int_{0}^{\infty}|C_{X,Y}(t)|\,dt
\end{equation}
In practice, the upper limit of the integral is replaced by a
value $t_{max}$ such that the correlation function can be
considered as decayed to its asymptotic value $C_{X,Y}=0$ ($t_{max}= 50$
for the data shown in figure \ref{fig3}). We have plotted these two
correlation times in the left panel of figure \ref{fig4}. Both times reach a maximum at approximately the same value $N\approx 160$, indicating that, for the set of parameters chosen, the maximum extent of the time correlation occurs for this number of coupled excitable systems.

Another common indicator for the regularity of the emitted pulses can be obtained by the {\sl jitter} of the time between pulses\cite{PK97}. A pulse in the $X(t)$ variable is defined when $X(t)$ exceeds a certain threshold
value $X_0$ (taken arbitrarily as $X_0=0.3$, although other values yield
similar results). The jitter $R_X$ is defined as the root mean square of the time $T_X$ between two consecutive pulses normalized to its mean value:
\begin{equation}
R_X =\frac{\sigma[T_X]}{\langle T_X\rangle}
\end{equation}
and an equivalent definition for the jitter $R_Y$ of the $Y$ variable.  The
smaller the value of $R_{X,Y}$, the larger the regularity of the pulses (a
value of $R_{X,Y}=0$ indicates a perfectly periodic signal). It is shown in the
right panel of figure \ref{fig4} that indeed the jitter in both variables have a
well defined minimum at a value of $N\approx 80$, again showing the existence
of the system size resonance.  When comparing with the results of the correlation time, it is not uncommon that the two indicators (the
correlation time $\tau$ and the jitter $R$) have their optimal values at
different values of the system parameters\cite{PK97,PTMCG01}.

In summary, we have shown that an ensemble of globally coupled FitzHugh--Nagumo
excitable systems subjected to independent noises pulse on average with a
regularity that is maximum for a given value of the number $N$ of coupled
systems. An approximate calculation indicates that the collective variable
$Y(t)$ is subjected to a noise of effective intensity $D/\sqrt{N}$. Therefore,
even in the presence of a large amount of noise ($D$ large), it is possible to
couple the right number of systems in order to optimize the periodicity of the
emitted pulses. Since the FitzHugh--Nagumo system has been used previously to
model some biological systems, we believe that our results can be relevant when
analyzing the collective response of such systems in a noisy environment and
can help to explain the observed size of some ensembles of excitable cells in
living organisms.

\noindent{\bf Acknowledgments:} This work is supported by the Ministerio de
Ciencia y Tecnolog{\'\i}a (Spain) and FEDER, projects BFM2001-0341-C02-01 and
BFM2000-1108, and NSF grant DMR9813409. J.D.G. acknowledges IBERDROLA for
financial support.

\newpage

\end{document}